# Equilibrium and Stability of Elliptical Galaxies[1]

D. Merritt and T. Fridman
*Department of Physics and Astronomy, Rutgers University, New Brunswick, NJ 08855*

**Abstract.** Recent work on the equilibrium and stability of ellipsoidal stellar systems is reviewed. The absence of constant-density cores in early-type galaxies implies that chaos and high-order resonances are generic features of the motion in triaxial stellar systems. Bending instabilities are now well understood and may provide the answers to a number of long-standing riddles, including the absence of highly flattened elliptical galaxies and the formation of galactic bulges. Some preliminary results on instabilities in two-integral oblate models are presented.

## 1. Triaxial Galaxies with Cusps

Lees & Schwarzschild (1992) divide nonsingular triaxial potentials into three radial sections which differ from each other in their orbital content. The central two sections are present only in potentials with a core; they extend out to the radius where the $x$-axial orbit, which generates the box orbits, becomes unstable. Inside of this radius the orbits are mostly regular, i.e. they respect three isolating integrals of the motion. Typically there are four major orbit families: the box orbits, which have filled centers; and the three families of tube orbits, which avoid the center. These four families are the same ones that appear in fully integrable, Stäckel potentials (Kuzmin 1973; de Zeeuw and Lynden-Bell 1985). Beyond a few tens of core radii (the exact radius depends on the model shape and density profile), the $x$-axial orbit becomes unstable and the box orbits disappear. Orbits which — like the boxes — touch an equipotential surface are now either stochastic, i.e. lack two of their three integrals of motion, or else they belong to a family of regular orbits associated with a stable resonance that avoids the center. The latter are called "boxlets" and the most important boxlets are those generated by the 1 : 2 "banana" orbits in the $x - z$ plane. [1]

Recent ground-based (Moller *et al.* 1995) and HST (Crane *et al.* 1993; Jaffe *et al.* 1994; Ferrarese *et al.* 1994; Lauer *et al.* 1995) observations reveal that early-type galaxies never have constant-density cores; the stellar surface brightness always continues to rise, roughly as a power law, into the smallest observable radius. If the mass distributions follow suit, then the orbital motion in elliptical galaxies may bear little resemblance to the motion in Stäckel potentials at *any* radius, either large or small.

---

[1] To appear in **Fresh Views of Elliptical Galaxies**, ed. A. Buzzoni, A. Renzini & A. Serrano, 1995.

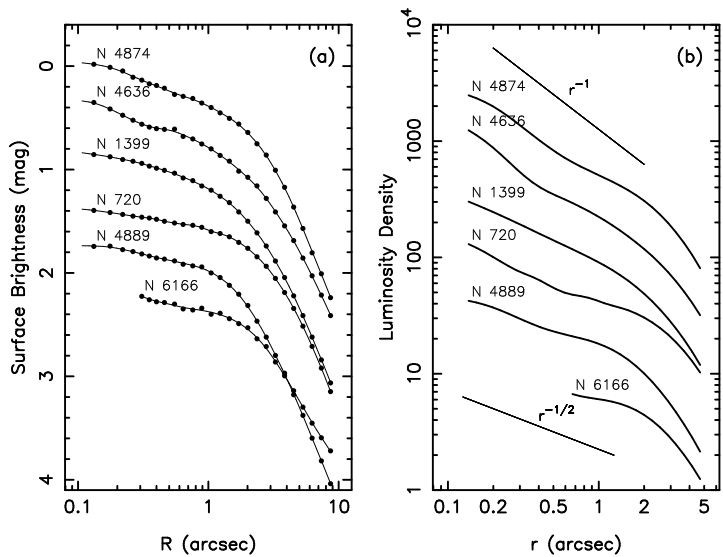

Figure 1. Surface brightness (a) and luminosity density (b) profiles for six elliptical galaxies observed with HST. The data points in (a) are from Lauer *et al.* (1995); solid curves are nonparametric fits.

Ferrarese *et al.* (1994) and Lauer *et al.* (1995) divide elliptical galaxies into two classes based on their nuclear properties, which Lauer *et al.* call "core" and "power-law" galaxies. Core galaxies exhibit a definite break in the surface brightness profile at some radius $R_b$; inward of this break, the profile turns down to a shallow inner power law $\Sigma(R) \propto R^{-\alpha}$, $\alpha \approx -0.1 \pm 0.1$. Power-law galaxies show essentially a single power-law profile throughout their inner regions with $\alpha \approx -0.8 \pm 0.2$. Power-law galaxies are of lower average luminosity than core galaxies, but steep surface brightness profiles are seen in galaxies with a wide range of absolute magnitudes, from $M_v \approx -16$ or fainter (e.g. M32) to $M_v \approx -21$ (e.g. NGC 4621) (Kormendy *et al.* 1995).

The luminosity density of a galaxy with a central, $\Sigma \propto R^{-1}$ cusp is approximately $\rho \propto r^{-2}$ at small radii; thus the power-law galaxies have a roughly inverse-square dependence of luminosity density on radius near their centers. No such simple conclusion can be drawn for the "core" galaxies, since a luminosity density that rises as $\rho \propto r^{-1}$ or more slowly near the center appears in projection to have a gently curving surface brightness profile, with a slope that falls to zero at the center (e.g. Dehnen 1993, Fig. 1). Thus the core galaxies could have density cusps as steep as $\rho \propto r^{-1}$.

Figure 1 shows model-independent estimates of the density profiles of six core galaxies, based on the surface brightness data in Table 3 of Lauer *et al.* (1995).

Smoothing splines (Wahba 1990) were fit to the surface brightness data and analytically deprojected; unlike parametric model fitting, this approach is "consistent," i.e. gives correct results in the limit of perfect data (Nychka et al. 1984). Figure 1 reveals that the core galaxies, like the power-law galaxies, do in fact have approximate power-law luminosity density profiles near their centers, with slopes that range from $\rho \propto r^{-0.5}$ to $\rho \propto r^{-1}$.

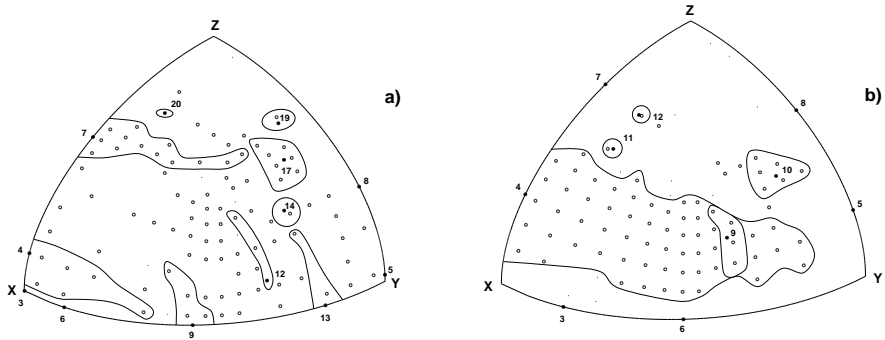

Figure 2. Stationary start space at one energy for two triaxial models with $r^{-1}$ (a) and $r^{-2}$ (b) central density cusps. Dots are regular orbits; numbered dots are resonant orbits.

The central force diverges for mass distributions steeper than $\rho \propto r^{-1}$, and the axial orbits in a triaxial model with a cusp like those in Lauer et al.'s power-law galaxies are unstable at all energies. Schwarzschild (1993) has cataloged the orbital motion in a set of six, scale-free triaxial potentials with $\rho \propto r^{-2}$. He found that orbits which begin on an equipotential surface are mostly stochastic, except for those associated with stable resonances like the bananas. Non-scale-free triaxial potentials with both $r^{-2}$ and $r^{-1}$ density cusps were investigated by Merritt & Fridman (1995). They found a similar fraction of stochastic orbits in the two potentials; the main difference was a greater number of stable resonant families at low energies in the potential with the weaker cusp (Figure 2).

Less is known about the motion in triaxial potentials with the still weaker cusps that are seen in Lauer et al.'s "core" galaxies. In a spherical model with Dehnen's (1993) density law,

$$\rho(r) = \frac{(3-\gamma)aM}{4\pi} r^{-\gamma}(a+r)^{-(4-\gamma)}, \qquad (1)$$

the radial force peaks at $r = a(1-\gamma)/2$ for $\gamma \leq 1$; thus, from a dynamical point of view, such models have "cores" even though the density diverges. We can define a "dynamical core radius" $r_d$ as the radius of peak radial force; this definition is consistent with, for instance, the usual definition of the core radius in an isothermal sphere. Based on Lees & Schwarzschild's (1992) arguments, we might expect the orbital motion in the core galaxies to be essentially regular only within a region extending out to some multiple of $r_d$, where the $x$-axial orbit first becomes unstable. Numerical integrations reveal that the $x$-axial orbit is in fact unstable at *all* energies in triaxial potentials when $\gamma \gtrsim 0.5$; setting $\gamma$ to zero gives instability at $\sim 5r_d$ and above. Thus we expect bona-fide box orbits to appear only in triaxial models with quite weak cusps, $\gamma \lesssim 0.5$, and even then they should be restricted to the central regions.

Figure 3. Three cross-sections through a regular boxlet (a), a time-averaged stochastic orbit (b) and a "fully ergodic" orbit that uniformly fills the energy surface (c) in a cuspy triaxial model with $\gamma = 2$. The number in each cell represents the fraction of time which the orbit spends in the cell.

These arguments suggest that chaos and high-order resonant families are generic features of the motion in realistic triaxial potentials. How should the stochastic orbits be included in a self-consistent solution? The answer is partly a matter of time scale. Regular orbits are quasi-periodic and ergodic: the motion can be represented as a sum of oscillations about a three-dimensional torus in phase space, and (except in the case of resonant orbits, for which the frequencies are commensurable) this motion eventually fills the surface of the torus densely and uniformly. Stochastic orbits are qualitatively different. Although they can not wander freely over the entire energy hypersurface – the parts of phase space associated with regular tori are excluded, for instance – it is believed that stochastic orbits densely fill a region called the "Arnold web" that links together all stochastic trajectories at a given energy. Thus, there is only one, time-averaged stochastic orbit at every energy, but the time required for a stochastic trajectory to fill the web in an effectively ergodic manner can be very long. Figure 3 shows the configuration-space density of a stochastic orbit that was integrated for $10^5$ dynamical times in a triaxial model with $\gamma = 2$. The density distribution is more nearly round than that of a typical boxlet, but more flattened than that of an orbit that wanders freely over the energy surface.

Numerical integrations suggest that the stochastic orbits are effectively ergodic over time scales of order $10^3$ to $10^4$ dynamical times in a triaxial model with a $\rho \propto r^{-2}$ density cusp (Merritt & Fridman 1995). In a model with a weaker $r^{-1}$ cusp, the fraction of stochastic phase space is nearly as large but the diffusion times are longer. Thus many of the stochastic orbits in a triaxial galaxy with a weak cusp would behave much like regular orbits over the lifetime of a galaxy.

A self-consistent model that included only long-time-averaged stochastic orbits would be fully stationary, i.e. it would satisfy a version of Jeans's theorem. A model that included stochastic orbits integrated for shorter times would be quasi-stationary at best, since the stochastic orbits would not have attained their time-averaged densities. Schwarzschild (1993) found that his scale-free triaxial models could sometimes not be built up from the regular orbits alone; however quasi-equilibria that included the stochastic orbits, integrated for only 55 dynamical times, could always be found. Merritt & Fridman (1995) attempted to build non-scale-free models in which the stochastic orbits were excluded, or time-averaged, at low energies where diffusion times are the shortest. These "fully mixed" solutions could be found for $\gamma = 1$, in the sense that a large fraction of the mass near the center of the model could be placed on time-averaged stochastic orbits without violating the self-consistency. However no significant fraction of the mass could be placed on fully-mixed stochastic orbits for $\gamma = 2$, demonstrating that triaxiality can be inconsistent with a strong central cusp, as first suggested by Gerhard & Binney (1985).

Even in galaxies with weak cusps, however, nature may not be inclined to populate the stochastic orbits in just the right way to guarantee full equilibrium, especially at large radii where diffusion time scales for stochastic orbits are long. Thus, slow evolution — toward sphericity, or at least axisymmetry — may be a common property of triaxial stellar systems.

## 2. Bending Instabilities

As important as studies of equilibrium models are for understanding the possible dynamical states of real galaxies, nature requires more than just equilibrium. A real stellar system must be stable to small perturbations as well. Here we discuss recent work on bending instabilities in hot stellar systems. These instabilities may be relevant to a number of long-standing puzzles in galactic dynamics, including the absence of highly elongated elliptical galaxies; the origin of bulges; and the fact that most disk galaxies do not contain strong bars.

Toomre (1966) pointed out many years ago that any sufficiently thin stellar system in which some component of the kinetic energy is in the form of random or counterstreaming motions will be susceptible to a bending instability. This instability has since been called the "hose" or "firehose" instability by analogy with a similar instability in magnetized plasmas (Parker 1958). A more suitable term is "bending instability", since it is far from clear that Toomre's instability has anything to do with fire hoses. (The mental picture is apparently that of a fire hose gyrating wildly as water spews from the unsupported nozzle.) A better analogy, as first pointed out by Toomre (1966), is with the Kelvin-Helmholtz instability that occurs when two fluids slide past one another, or with beads sliding along an oscillating string (Parker 1958).

Toomre (1966) showed that a thin stellar sheet has a tendency to buckle when the stars have large random velocities along the plane and are constrained to remain in a single thin layer as the sheet bends. The constraining force comes from the vertical self-gravity of the sheet, which is large if the sheet is thin. Stars moving across a bend are forced to oscillate vertically as they pursue their unperturbed horizontal motions, and the bend will grow if the gravitational restoring forces from the perturbation are too weak to provide the vertical acceleration required.

Toomre's thin-sheet dispersion relation is

$$\omega^2 = 2\pi G \Sigma k - \sigma_u^2 k^2. \qquad (2)$$

The first term, which arises from the perturbed gravity in a sheet of surface density $\Sigma$, is stabilizing, while the second term, due to the centrifugal force that stars with velocity dispersion $\sigma_u = \langle u^2 \rangle^{1/2}$ exert on the sheet, is destabilizing. For sufficiently long wavelengths $\lambda > \lambda_J = \sigma_u^2/G\Sigma$, the gravitational restoring force dominates and the sheet is stable. Bending instabilities are precisely complementary, in this sense, to the Jeans instability in the plane, which is stabilized at wavelengths $\lambda < \lambda_J$.

Toomre's dispersion relation demonstrates that any perfectly thin, hot system will be unstable to bending at short wavelengths. At wavelengths shorter than the actual vertical thickness of the system, however, we would expect the bending to once again be stabilized. The reason is that stars in a finite-thickness system oscillate vertically with an unperturbed frequency $\kappa_z$; like any oscillator, the phase of the star's response to the imposed bending depends entirely on whether the forcing frequency $ku$ is greater than or less than its natural frequency. If $ku > \kappa_z$ for most stars, the overall density response to the perturbation will produce a potential opposite to that imposed by the bend and the disturbance will be damped (Merritt & Sellwood 1994). These arguments suggest that a sufficiently thick system with low $\kappa_z$ will be stable to bending at all wavelengths, both short and long.

Araki (1985) derived the exact linear normal modes of a finite-thickness slab with anisotropic Gaussian velocity distributions. He found that bending at all wavelengths was stabilized when the ratio of vertical to horizontal velocity dispersions exceeded $\sim 0.293$, confirming a (remarkably accurate) earlier estimate of 0.3 made by Toomre (1966). Since the elongation of a pressure-supported galaxy with this anisotropy is approximately 15 : 1, Araki's result suggested that bending modes were unlikely to be of much importance in elliptical galaxies. However Fridman & Polyachenko (1984) showed that the critical axis ratio for stability of homogeneous oblate and prolate models was roughly 3 : 1, not 15 : 1, and Merritt & Hernquist (1991) found a similar result in their $N$-body study of inhomogeneous prolate spheroids.

The discrepancy was explained by Merritt & Sellwood (1994), who showed that the gravitational restoring force from the bend is substantially weaker in finite or inhomogeneous models than in infinite sheets and slabs. In fact the long-wavelength modes in thin systems with realistic density profiles are often not stabilized by gravity; this is especially true when the velocity dispersion is high in the direction of the bend, as would be the case in a prolate galaxy that bends around a short axis, for instance. In such a system, a typical star feels a vertical forcing frequency from a long-wavelength bend that is roughly twice the frequency $\Omega_x$ of its orbital motion along the long axis. Stability to global bending modes then requires that this forcing frequency be greater than $\Omega_z$, the frequency of orbital motion parallel to the short axis. The resulting condition $2\Omega_x > \Omega_z$ predicts stability for pressure-supported, homogeneous prolate spheroids rounder than 2.94 : 1, in excellent agreement with the normal-mode calculations of Vandervoort (1991) and Fridman & Polyachenko (1984), and consistent with Merritt & Hernquist's (1991) $N$-body study of inhomogeneous prolate models.

The situation for oblate models is more complicated since the shapes of the dominant modes depend on whether the velocities are azimuthally or radially biased. In oblate models with radially-elongated velocity ellipsoids, arguments similar to those given above suggest that an axis ratio of roughly 3 : 1 is again close to critical, in agreement with the $N$-body results for thickened disks (Sellwood & Merritt 1994). If the velocities are azimuthally biased, the orbits are approximately circular and so the dominant modes would be angular ones, $\delta z \propto e^{im\phi}$. The approximate condition for stability becomes $m\Omega > \kappa_z$, with $\Omega$ the circular orbital frequency.

A nice set of models with which to check this prediction are the scale-free, nonrotating, two-integral oblate models of Toomre (1982). Toomre's models have a velocity distribution function

$$f(E, L_z) = f_0 L_z^{2n} e^{-E/\sigma^2} \tag{3}$$

and a density

$$\rho(r, \theta) = \rho_0 S_n(\theta)(r_0/r)^2 \tag{4}$$

with $S_n(\theta)$ given by Toomre's equation (21). The model flattening increases with $n$; $n = 0$ is spherical and the $n = 3$ model has an average flattening of about 5 : 1.

Any two-integral model becomes radially "cold" as the flattening is made extreme, by virtue of $\sigma_R = \sigma_z$. Thus for high $n$ the dominant motion in Toomre's models is circular. Near the equatorial plane, these models have $\kappa_z^2 = (n + 1)V_0^2/R^2$, $V_0^2 = 4\pi G\rho_0 r_0^2$. According to the criterion given above, stability to bending modes of the form $e^{im\phi}$ requires $\kappa_z < m\Omega = mV_0/R$, or $m^2 > 1 + n$. Thus the $n = 3$ model should just be stable to $m = 2$ (saddle) modes; the $n = 8$ model to $m = 3$ modes; etc.

Recently, A. Toomre (unpublished) has computed the exact linear response of his models to bending modes. For $m = 2$, he finds $n = 3.124$ to be the critical value of flattening — satisfyingly close to $n = 3$. This example, and others given in Merritt & Sellwood (1994), suggest that the physics of bending modes in stellar systems are now pretty well understood.

Fridman & Polyachenko (1984) were the first to suggest that the absence of elliptical galaxies flatter than $\sim$ E6 was due to bending instabilities. The work summarized above greatly strengthens their hypothesis, at least to the extent that the stellar motions in elliptical galaxies are mostly "in and out" as opposed to "around and around." Oblate galaxies with nearly circular orbits — like those in flattened, two-integral models — can remain stable to bending modes even when somewhat flatter, e.g. $\sim$ 5 : 1 in the models just discussed. If more than half the stars in an oblate model are set to orbiting in the same direction, producing a net rotation, the critical flattening can be even greater; even perfectly thin disks are stable to bending modes when all the stars orbit in one direction (Hunter & Toomre 1969), since the bending wave is free to travel with the particles.

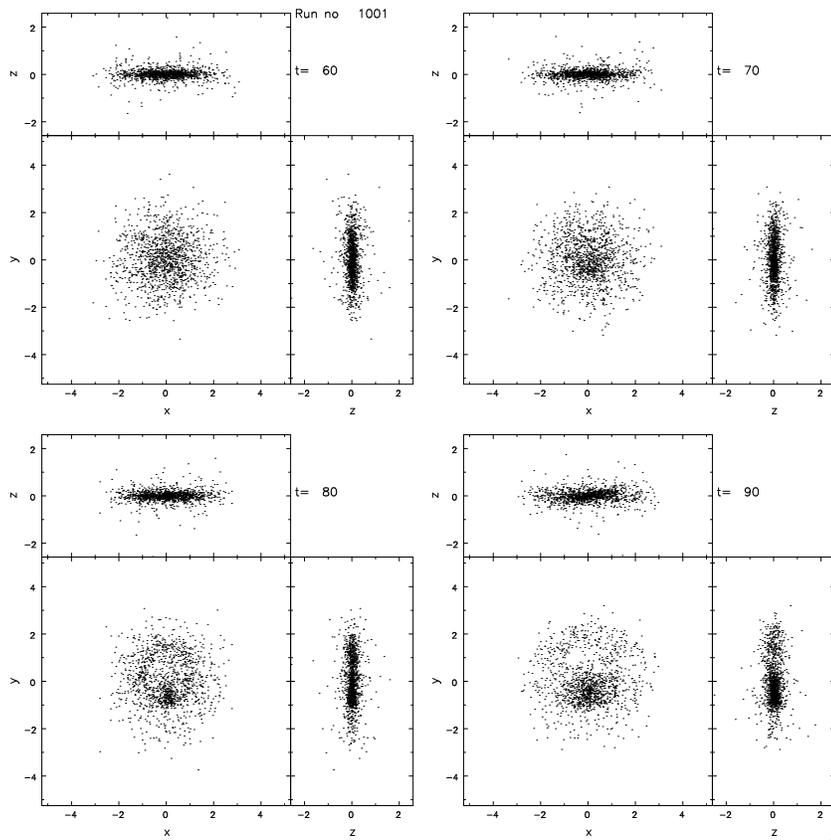

Figure 4. Lopsided instability in a nonrotating oblate model from the Kuzmin-Kutuzov (1962) family.

Although strongly counterrotating, flattened stellar systems may seem contrived, nature apparently does make them. The best-studied example is NGC 4550, an S0 galaxy that consists of two, nearly equal, counterrotating streams of stars (Rubin *et al.* 1992). A number of other examples of strongly counterrotating galaxies have since been found (Merrifield & Kuijken 1994). Numerical experiments have successfully produced stable, flattened oblate models that look quite similar to these galaxies (Sellwood & Merritt 1994).

A more exciting, but still rather speculative, application of these bending modes is to the formation of bulges in disk galaxies. Rapidly-rotating disks are unstable to the formation of bars; this instability is an embarrassment to galactic dynamicists since most disk galaxies do not contain a strong bar. Early indications that these bars might thicken with time (e.g. Combes *et al.* 1990) were spectacularly confirmed by Raha *et al.* (1991), who carried out three-dimensional simulations which revealed a violent buckling instability. This is undoubtedly the same instability described above, driven in this case by the largely in-and-out motion along the long axis of the thin bar. The result is a thick peanut-shaped system that looks very much like the bulges in some edge-on disk galaxies (Shaw 1987). Seen face-on, the bar appears to be weakened or even destroyed by the instability; this may account for the low fraction of disk galaxies observed to contain strong bars.

## 3. Lopsided Instabilities in Oblate Models

Two-integral oblate models, $f = f(E, L_z)$, are now much in vogue for modelling real elliptical galaxies (de Zeeuw, this volume). Until recently, the stability of such models had never been systematically checked. Here we report preliminary results from an $N$-body study of the stability of two families of nonrotating, oblate models with $f = f(E, L_z)$ (Valluri, Merritt & Sellwood, in preparation).

At least two types of unstable evolution are known to be important in nonrotating oblate models when their flattening is sufficiently extreme. The corrugation modes described above should appear when the ellipticity exceeds 5 : 1 or so. In addition, lopsided, in-plane modes are known to exist in radially cold oblate models regardless of their flattening (Merritt & Stiavelli 1990). Since two-integral models become radially cold as they are made flat, we would expect these lopsided modes to appear in any family of two-integral models when their elongation exceeds some critical value.

Figure 4 shows the growth of an unstable lopsided mode in an oblate model from the Kuzmin & Kutuzov (1962) family. These modes were found to be present in oblate models with elongations of $\sim$ 5 : 1 or more; their net effect is to increase the radial pressure without greatly modifying the mass distribution. Thus, while bending modes place limits on the flattening of two-integral oblate models, the lopsided modes limit the fraction of kinetic energy that can assigned to nearly-circular motions. However both modes appear to be important only in highly flattened models.

The dynamical mechanism underlying these lopsided modes is not well understood and is a fruitful topic for further study.
identification

**Acknowledgments.** The $N$-body results shown in Figure 4 were prepared by Monica Valluri. This work was supported by NSF grant AST 90-16515 and NASA grant NAG 5-2803.

**Discussion**

*T. de Zeeuw*: Your numerical experiments suggest that it is difficult to construct triaxial galaxies with cusps. You have however tried the hardest case, namely the shape where the tube orbits occupy the smallest fraction of phase space. Your results may suggest that triaxial cusps exist only for certain near-oblate or near-prolate shapes, where the tube orbits are more dominant.

*D. Merritt*: I agree.

*A. Renzini*: You mentioned bending instabilities as a way of making bulges. How would the initial conditions (nearly equal numbers of rotating and counter-rotating stars in a thin disk) be established in the first place?

*D. Merritt*: The idea is that a bar forms first, through the usual instability of a rapidly-rotating disk. The stellar motions in a bar are largely counter-streaming, up and down along the bar's long axis. This motion, coupled with the thinness of the bar, drives the subsequent bending instability.